\pdfoutput=1
\documentclass[twocolumn, tighten]{myaastex62}
\usepackage{amsmath}
\shorttitle{Spaceflight from Super-Earths is difficult}
\shortauthors{Michael Hippke}
\begin{document}
\title{SPACEFLIGHT FROM SUPER-EARTHS IS DIFFICULT}

\author[0000-0002-0794-6339]{Michael Hippke}
\affiliation{Sonneberg Observatory, Sternwartestr. 32, 96515 Sonneberg, Germany}
\email{michael@hippke.org}

\begin{abstract}
Many rocky exoplanets are heavier and larger than the Earth, and have higher surface gravity. This makes space-flight on these worlds very challenging, because the required fuel mass for a given payload is an exponential function of planetary surface gravity. We find that chemical rockets still allow for escape velocities on Super-Earths up to $10\times$ Earth mass. More massive rocky worlds, if they exist, would require other means to leave the planet, such as nuclear propulsion.\\
\end{abstract}

\section{Introduction}
Do we inhabit the best of all possible worlds \citep{Leibnitz1710}? From a variety of habitable worlds that may exist, Earth might well turn out as one that is marginally habitable. Other, more habitable (``superhabitable'') worlds might exist \citep{2014AsBio..14...50H}. Planets more massive than Earth can have a higher surface gravity, which can hold a thicker atmosphere, and thus better shielding for life on the surface against harmful cosmic rays. Increased surface erosion and flatter topography could result in an ``archipelago planet'' of shallow oceans ideally suited for biodiversity. There is apparently no limit for habitability as a function of surface gravity as such \citep{2017arXiv171005605D}. Size limits arise from the transition between Terran and Neptunian worlds around $2\pm0.6\,R_{\oplus}$ \citep{2017ApJ...834...17C}. The largest rocky planets known so far are $\sim1.87\,R_{\oplus}$, $\sim9.7\,M_{\oplus}$  \citep[Kepler-20\,b,][]{2016AJ....152..160B}. When such planets are in the habitable zone, they may be inhabited. Can ``Super-Earthlings'' still use chemical rockets to leave their planet? This question is relevant for SETI and space colonization \citep{2016AsBio..16..418L,2016arXiv160808770F,2017arXiv170703730F}.

\section{Method}
At our current technological level, spaceflight requires a rocket launch to provide the thrust needed to overcome Earth's force of gravity. Chemical rockets are powered by exothermic reactions of the propellant, such as hydrogen and oxygen. Other propulsion technologies with high specific impulses exist, such as nuclear thermal rockets \citep[e.g., NERVA,][]{1969JSpRo...6..565A}, but have been abandoned due to political issues. Rockets suffer from the \citet{tsiolkovsky1903issledovanie} equation : if a rocket carries its own fuel, the ratio of total rocket mass versus final velocity is an exponential function, making high speeds (or heavy payloads) increasingly expensive \citep{1992CeMDA..53..227P}.

The achievable maximum velocity change of a chemical rocket is

\begin{equation}
    \Delta v = v_\text{ex} \ln \frac{m_0}{m_f}
\end{equation}

where $m_{0}$ is the initial total mass (including fuel), $m_{f}$ is the final total mass without fuel (the dry mass), and $v_{\rm ex}$ is the exhaust velocity. We can substitute $v_{\rm ex}=g_{\rm 0}\,I_{\rm sp}$ where $g_{\rm 0}=G\,M_{\oplus}/R^2_{\oplus}\sim9.81\,$m\,s$^{-2}$ is the standard gravity and $I_{\rm sp}$ is the specific impulse (total impulse per unit of propellant), typically $\sim350\dots450\,$s for hydrogen/oxygen.

To leave Earth's gravitational influence, a rocket needs to achieve at minimum the escape velocity

\begin{equation}
    v_{\rm esc} = \sqrt{\frac{2GM_{\oplus}}{R_{\oplus}}}\sim11.2\,{\rm km\,s}^{-1}
\end{equation}

for Earth, and $v_{\rm esc}\sim27.1\,$km\,s$^{-1}$ for a $10\, M_{\oplus}$, $1.7\,R_{\oplus}$ Super-Earth similar to Kepler-20\,b.

\section{Results}
We consider a single-stage rocket with $I_{\rm sp}=350\,$s and wish to achieve $\Delta v > v_{\rm esc}$. The mass ratio of the vehicle becomes

\begin{equation}
    \frac{m_0}{m_f} > {\rm exp} \left( \frac{v_{\rm esc}}{v_\text{ex}} \right).
\end{equation}

which evaluates to a mass ratio of $\sim26$ on Earth, and $\sim2{,}700$ on Kepler-20\,b. Consequently, a single-stage rocket on Kepler-20\,b must burn $104\times$ as much fuel for the same payload ($\sim2{,}700$\,t of fuel for each t of payload).

This example neglects the weight of the rocket structure itself, and is therefore a never achievable lower limit. In reality, rockets are multistage, and have typical mass ratios (to Earth escape velocity) of $50\dots150$. For example, the Saturn~V had a total weight of 3{,}050\,t for a lunar payload of 45\,t, so that the ratio is 68. The Falcon Heavy has a total weight of 1{,}400\,t and a payload of 16.8\,t, so that the ratio is 83 (i.e., the payload fraction is $\sim1\,$\%).

For a mass ratio of 83, the minimum rocket (1\,t to $v_{\rm esc}$) would carry $9{,}000\,$t of fuel on Kepler-20\,b, which is $3\times$ larger than a Saturn~V (which lifted 45\,t). To lift a more useful payload of 6.2\,t as required for the James Webb Space Telescope on Kepler-20\,b, the fuel mass would increase to $55{,}000$\,t, about the mass of the largest ocean battleships. For a classical Apollo moon mission (45\,t), the rocket would need to be considerably larger, $\sim400{,}000$\,t. This is of order the mass of the Pyramid of Cheops, and is probably a realistic limit for chemical rockets regarding cost constraints.

\section{Discussion}

\subsection{Launching from a mountain top}
Rockets work better in space than in an atmosphere. One might consider launching the rocket from high mountains on Super-Earths. The rocket thrust is given by

\begin{equation}
F = \dot{m}\,v_{ex} + A_e(P_1 - P_2)
\end{equation}

where $\dot{m}$ is the mass flow rate, $A_e$ is the cross-sectional area of the exhaust jet, $P_1$ is the static pressure inside the engine, and $P_2$ is the atmospheric pressure. The exhaust velocity is maximized for zero atmospheric pressure, i.e. in vacuum. Unfortunately, the effect is not very large in practice. For the Space Shuttle's main engine, the difference between sea level and vacuum is $\sim25\,$\% \citep{Rocketdyne}. Atmospheric pressure below 0.4\,bar (Earth altitude $6{,}000$\,m) is not survivable long term for humans, and presumably neither for ``Super-Earthlings''. Such low pressures are reached in lower heights on Super-Earths, because the gravity pulls the air down.

One disadvantage is that the bigger something is, the less it can deviate from being smooth. Tall mountains will crush under their own weight \citep[the ``potato radius'' is $\sim238\,$km,][]{2015arXiv151104297C}. The largest mountains in our solar system are on less massive bodies, such as the Rheasilvia central peak on Vesta (22\,km) or Olympus Mons on Mars (21.9\,km). Therefore, we expect more massive planets to have smaller mountains. This will be detectable through transit observations in future telescopes \citep{2018MNRAS.tmp..142M}. One option would be to build artificial mountains as launch platforms.

\subsection{Launching rockets from water-worlds}
Many habitable (and presumably inhabited) planets might be waterworlds \citep{2017MNRAS.468.2803S}, and intelligent life in water and sub-surface is plausible \citep{2017arXiv171109908L}. Can rockets be launched from such planets? We here neglect how chemical fuels, and whole rockets, are assembled on such worlds.

Rockets on waterworlds could either be launched from floating pontoon-based structures, or directly out of the water. Underwater submarine rocket launches use classical explosives to flash-vaporize water into steam. The pressure of the expanding gas drives the missile upwards in a tube. This works well for ICBMs launched from submerged submarines

These aquatic launch complications make the theory of oceanic rocket launches appear at first quite alien; presumably land-based launches seem equally human to alien rocket scientists.

\section{Conclusion}
The amount of fuel required per payload to escape velocity scales as $\sim 3.3\,\exp(g_{\rm 0})$. Chemical rocket launches are still plausible for Super-Earths $\lesssim10\,M_{\oplus}$, but become unrealistic for more massive planets. On worlds with a surface gravity of $\gtrsim 10\,g_{\rm 0}$, a sizable fraction of the planet would need to be used up as chemical fuel per launch, limiting the total number of flights.

As an alternative, space elevators may be considered. One limiting factor is tensile strength. The most suitable material known today, carbon nanotubes, is just barely sufficient for Earth's gravity \citep{2000AcAau..47..735E,2006JPCM...18S1971P}. It is unclear if stronger materials are physically possible. If this is not the case, space elevators on Super-Earths would not work. To our knowledge, the only option would then be nuclear-powered rockets.

\bibliography{references}
\end{document}